# NIR-to-NIR ratiometric and lifetime based luminescence thermometer on a structural phase transition in $Na_3Sc_2(PO_4)_3:Yb^{3+}$

Anam Javaid[1], Maja Szymczak[1], Malgorzata Kubicka[1], Justyna Zeler[2], Vasyl Kinzhybalo[1], Marek Drozd[1], Damian Szymanski[1], Lukasz Marciniak[1*]

[1] Institute of Low Temperature and Structure Research, Polish Academy of Sciences, Okólna 2, 50-422 Wrocław, Poland

[2]Faculty of Chemistry, University of Wroclaw, 14 F. Joliot-Curie Street, Wroclaw, 50-383 Poland

*corresponding author:

**Abstract**

The ratiometric approach is the most commonly employed readout technique in luminescence thermometry. To address the trade-off between the risk of measurement disturbance in thermometers with high spectral separation of emission bands (due to dispersion in the surrounding medium) and the low sensitivity observed in ratiometric thermometers based on Stark level thermalization, we propose a thermometer based on the structural phase transition in $Na_3Sc_2(PO_4)_3:Yb^{3+}$. The use of $Yb^{3+}$ ions as dopants and the changes in Stark level energies associated with the thermally induced monoclinic-to-trigonal phase transition enable the development of a thermometer with high relative sensitivity, achieving $S_{Rmax}$=1.5% $K^{-1}$ at 340K for $Na_3Sc_2(PO_4)_3$:0.1%$Yb^{3+}$. Additionally, as demonstrated, the structural transition alters the probability of radiative depopulation of the $^2F_{5/2}$ state of $Yb^{3+}$, allowing the development of a lifetime-based luminescence thermometer on $Na_3Sc_2(PO_4)_3$:$Yb^{3+}$ with $S_{Rmax}$= 1.2 %K$^{-1}$ at 355K for $Na_3Sc_2(PO_4)_3$:1%$Yb^{3+}$. Furthermore, the phase transition temperature and consequently the thermometric performance of $Na_3Sc_2(PO_4)_3$:$Yb^{3+}$ can be modulated by varying the $Yb^{3+}$ ion concentration, offering additional tunability for specific applications.

**Introduction**

The remarkable popularity of luminescence thermometry, which utilizes thermally induced changes in the spectroscopic properties of a phosphor to measure temperature, stems from several significant advantages over other remote and contact-based temperature measurement techniques[1–4]. Foremost among these is its ability to provide both spot temperature readings and two-dimensional imaging of temperature variations[5–8]. Additionally, luminescence thermometry offers high reliability, fast response times, and electrically passive operation, making it highly desirable for a wide range of applications[9,10]. Among the various thermally dependent spectroscopic parameters of phosphors, ratiometric and lifetime-based approaches dominate in popularity, as evidenced by the extensive number of published studies[11]. This preference stems from their superior reliability, precision, and repeatability[11,12]. Ratiometric temperature measurement, which relies on monitoring the temperature-dependent intensity ratio of two emission bands, is particularly advantageous due to its cost-effectiveness, as it does not require expensive equipment and can be easily implemented using affordable spectrometers or even digital cameras[8,13]. While many ratiometric luminescent thermometers have been proposed, the dispersive dependence of the extinction coefficient of the medium surrounding the thermometer or in the optical path to the detection system can compromise measurement reliability in certain cases[11]. Consequently, from this perspective it is highly desired for the spectral bands involved in temperature determination to be in close spectral proximity[11]. Although carefully selected pairs of ions emitting in nearby spectral regions can partially address these challenges, single-band ratiometric thermometers demonstrate exceptional thermometric performance[14–19]. Two main approaches to single-band ratiometric thermometers have been developed so far: (i) a method where a change in temperature induces a variation in the emission intensity ratio of two spectral bands under single-wavelength excitation[14–19], and (ii) a method where the thermometric parameter is the ratio of single emission band intensities under two different excitation wavelengths[20–22]. While later approach has demonstrated extremely high thermal sensitivity[20–22], the former approach is undoubtedly simpler to implement. In approach (i), the thermalization of Stark levels of $Ln^{3+}$ ions is typically exploited, where temperature changes alter the shape of the emission band[14,17,19]. However, the relatively small energy separations between Stark levels limit the achievable sensitivities.

To address these limitations, this study proposes the use of $Na_3Sc_2(PO_4)_3$ material doped with $Yb^{3+}$ ions as a luminescent thermometer. In $Na_3Sc_2(PO_4)_3:Yb^{3+}$, an increase in temperature induces a structural phase transition from a monoclinic low-temperature phase (LT) to a trigonal high-temperature phase (HT) [23–27]. Consequently, this phase transition significantly alters the local symmetry of $Yb^{3+}$ ions, modifying the splitting energies of the $^2F_{5/2}$ and $^2F_{7/2}$ Stark levels. As a result, the spectral positions of the emission lines shift, enabling ratiometric temperature measurement. Additionally, the structural changes associated with the phase transition impact the probability of radiative depopulation of the $^2F_{5/2}$ level, enabling the implementation of a lifetime-based thermometric approach in $Na_3Sc_2(PO_4)_3:Yb^{3+}$. Since the phase transition temperature is influenced by the composition of the host material, it is demonstrated that adjusting the concentration of $Yb^{3+}$ ions allows fine-tuning of the thermometric performance of this luminescent thermometer.

**Experimental Section**

*Synthesis*

A series of powder samples of $Na_3Sc_2(PO_4)_3$:x%$Yb^{3+}$, where x = 0.1, 1, 2, 5, 10, 15 and 20% were synthesized using a conventional high-temperature solid-state reaction method. $Na_2CO_3$ (99.9% of purity, Alfa Aesar), $Sc_2O_3$ (99.9% of purity, Alfa Aesar), $NH_4H_2PO_4$ (99.9% of purity, POL-AURA), and $Yb_2O_3$ (99.9% of purity, Stanford Materials Corporation) were used as starting materials. The raw materials were calculated based on the stoichiometric ratio, precisely weighed and finely ground in an agate mortar to achieve a homogeneous mixture. The mixture was subsequently transferred to an alumina crucible and calcined in air at 1300ºC for 6 hours (with a heating rate of 10 K min$^{-1}$). The final product was gradually cooled to the room temperature and then regrounded to obtain powder samples for structural and optical characterization.

*Methods*

The obtained materials were examined using powder X-ray diffraction technique. Powder diffraction data were obtained in Bragg–Brentano geometry using a PANalytical X'Pert Pro diffractometer equipped with Oxford Cryosystems Phenix (low-temperature measurements) attachment using Ni-filtered Cu K$\alpha$ radiation (V=40 kV, I=30 mA). Diffraction patterns in 2θ range of 15-90º were measured in cooling/heating sequence in the temperature range from 320 to 80 K. ICSD database entries No. 56865 (LT phase) and 65407 (HT phase) were taken as initial models for the analysis of the obtained diffraction data.

Morphology and chemical composition of the studied samples were determined with a Field Emission Scanning Electron Microscope (FE-SEM, FEI Nova NanoSEM 230) equipped with an energy-dispersive X-ray spectrometer (EDX, EDAX Apollo X Silicon Drift Correction) compatible with Genesis EDAX microanalysis Software. Before SEM imaging, the $Na_3Sc_2(PO_4)_3$:0.2%$Eu^{3+}$ sample (as a representative sample in the entire study series) was dispersed in alcohol, and then a drop of suspension was placed in the carbon stub. Finally, SEM images were recorded with an accelerating voltage of 5.0 kV in a beam deceleration mode which improves imaging parameters such as resolution and contras as well as reduces contamination. In the case of EDS measurements the sample was scanned at 30 kV.

A differential scanning calorimetric (DSC) measurements were performed using Perkin-Elmer DSC 8000 calorimeter equipped with Controlled Liquid Nitrogen Accessory LN2 with a heating/cooling rate of 20 K min$^{-1}$. The sample was sealed in the aluminum pan. The measurement was performed for the powder sample in the 100 - 800 K temperature range. The excitation and emission spectra were obtained using the FLS1000 Fluorescence Spectrometer from Edinburgh Instruments equipped with 450 W Xenon lamp and R928 photomultiplier tube from Hamamatsu. Luminescence decay profiles were also recorded using the FLS1000 equipped with 150 W µFlash lamp. The average lifetime ($\tau_{avr}$, Eq. 1) of the excited levels was calculated based on fit of the luminescence decay profiles by double-exponential function (Eq. 2):

$$\tau_{avr} = \frac{A_1\tau_1^2 + A_2\tau_2^2}{A_1\tau_1 + A_2\tau_2} \qquad (1)$$

$$I(t) = I_0 + A_1 \cdot \exp\left(-\frac{t}{\tau_1}\right) + A_2 \cdot \exp\left(-\frac{t}{\tau_2}\right) \qquad (2)$$

where $\tau_1$ and $\tau_2$ represent the luminescence decay parameters and $A_1$, $A_2$ are the fitted amplitudes of the double-exponential function. During the temperature-dependent emission measurements, the temperature of the sample was controlled by a THMS600 heating-cooling stage from Linkam (0.1 K temperature stability and 0.1 K set point resolution).

**Results and discussion**

$Na_3Sc_2(PO_4)_3$ is an intriguing material due to its structural properties, as it can exist in three distinct phases: monoclinic $α$ and trigonal $β$ and $γ$[23,24,26–30]. The formation of these different phases is influenced by various factors, ranging from synthesis conditions such as temperature to the type and amount of cation dopant in the structure. As demonstrated by Liu et al.[24] even a relatively small amount of $Eu^{2+}$ ions as dopants can lead to the formation of a trigonal structure under identical synthesis conditions that otherwise yield a monoclinic structure in the case of un-doped counterpart. From the perspective of this research, the most critical factor determining the structure of $Na_3Sc_2(PO_4)_3$ is temperature. Temperature can induce both irreversible structural changes, through alterations in synthesis conditions, and reversible structural phase transitions. Regarding temperature-induced first order phase transition, the literature reports two possible transitions: (1) a transition from the monoclinic $α$-phase to the trigonal $β$-phase at approximately 320 K, and (2) a transition from the $β$-phase to the $γ$-phase - both of which are trigonal - at approximately 440 K. The term "approximately" is used here due to the lack of consistent data on transition temperatures, as these depend on factors such as synthesis conditions and particles size, both of which influence the phase transition temperatures. Both the trigonal and monoclinic structures of $Na_3Sc_2(PO_4)_3$ are composed of octahedrally coordinated $Sc^{3+}$ ions, $Na^+$ ions in 5-, 7-, or 8-fold coordination, and phosphate groups ($PO_4^{3-}$) where $P^{5+}$ is tetrahedrally coordinated by $O^{2-}$. This visualization of the structures is schematically illustrated in Figure 1a. In terms of dopant ion substitution, the most likely positions for $Yb^{3+}$ ions within the $Na_3Sc_2(PO_4)_3$ structure are the crystallographic sites of $Sc^{3+}$ ions. This preference is primarily due to their identical ionic charge, which minimizes the need for charge compensation. Additionally, the ionic radii of $Sc^{3+}$ (0.745 Å) and $Yb^{3+}$ (0.868 Å) are quite similar, differing by only about 14%. This close match in ionic radii strongly supports the feasibility of substitution. The room temperature XRD patterns of $Na_3Sc_2(PO_4)_3:Yb^{3+}$ with different concentration of $Yb^{3+}$ ions correlates well with the reference pattern (Figure 1b, see also Figure S1). However, the careful analysis of these patterns indicates that with increase of $Yb^{3+}$ concentration the contribution of the HT phases gradually increases from 0.2% for 0.1% $Yb^{3+}$ to around 8% for 15% $Yb^{3+}$ (Figure 1c). This effect results from the lowering of the phase transition temperature with enlargement of the $Yb^{3+}$ ions concentration. To trace this thermally induced phase transition the XRD patterns of $Na_3Sc_2(PO_4)_3$:1% $Yb^{3+}$ were measured as a function of temperature. The Rietveld refinement analysis indicates (Figures S2-S30) that an increase in temperature above 300K results in an increase in HT phase contribution up to around 380K at which the sample consists only of HT phase (Figure 1d). The difference in the ionic radii between $Yb^{3+}$ dopant ions and $Sc^{3+}$ host material ions results in a gradual reduction of the temperature of the phase transition with increase in $Yb^{3+}$ concentration from 349.87 K for 0.1% $Yb^{3+}$ to 341. 47 K for 15% of $Yb^{3+}$ ions as confirmed by DSC studies (Figure 1e). The morphology of the examined samples were investigated by using SEM. Figure 1f shows the representative SEM image of $Na_3Sc_2(PO_4)_3$:0.1% $Yb^{3+}$ sample. Other $Yb^{3+}$-doped samples exhibit similar morphology characteristics, therefore their SEM images have not been presented here. As shown in Figure 1f, the $Yb^{3+}$-doped $Na_3Sc_2(PO_4)_3$ sample possess a typical morphology for materials synthesized by solid state method i.e. the irregular particles have from 5 to 10 microns in size which exhibit tend to aggregate into irregular clusters shapes up a few microns in size. From the elemental distribution maps of Na, P, Sc and Yb it is observable that above mentioned elements are uniformly distributed throughout the $Na_3Sc_2(PO_4)_3$:0.1% $Yb^{3+}$ sample. EDS measurements have also confirmed that samples were free from contaminants (Figures 1 g-j).

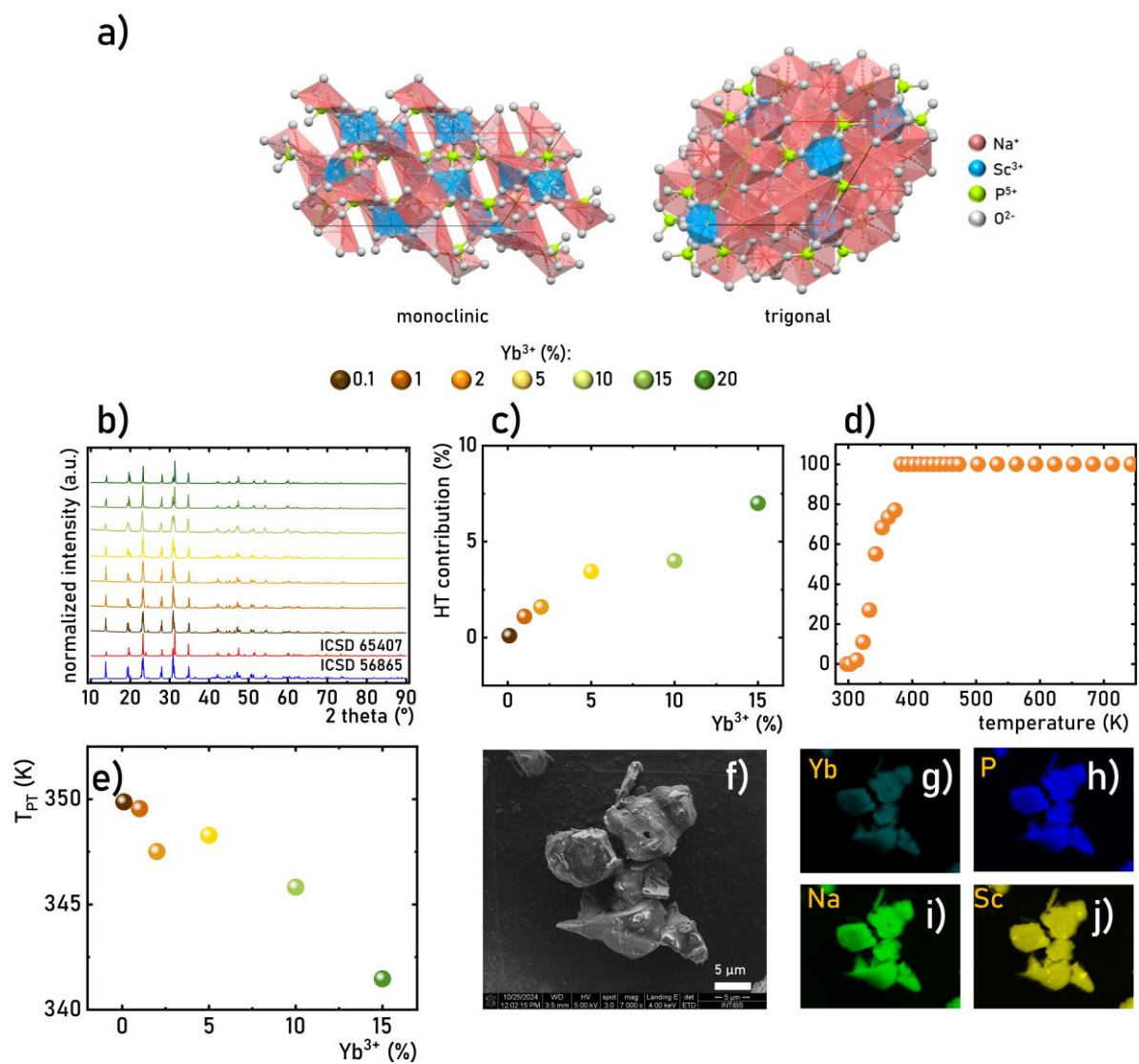

**Figure 1**. Visualization of the structure of the monoclinic and trigonal phases of $Na_3Sc_2(PO_4)_3$-a) the comparison of room temperature XRD patterns of $Na_3Sc_2(PO_4)_3$:$Yb^{3+}$ with different $Yb^{3+}$ concentration-b); the contribution of the trigonal phase of $Na_3Sc_2(PO_4)_3$ in the room temperature XRD patterns as a function of $Yb^{3+}$ concentration-c); the contribution of the trigonal phase of $Na_3Sc_2(PO_4)_3$ as a function of temperature for $Na_3Sc_2(PO_4)_3$:1% $Yb^{3+}$-d); temperature of the $\alpha \rightarrow \beta$ phase transition of $Na_3Sc_2(PO_4)_3$: $Yb^{3+}$ as a function of $Yb^{3+}$ concentration determined from the DSC analysis-e); representative SEM image -f) and corresponding elemental maps of the Yb (cyan), P (blue), Sc (yellow) and Na (green) for $Na_3Sc_2(PO_4)_3$:1% $Yb^{3+}$.

The Yb$^{3+}$ ions are distinguished by one of the simplest energy levels diagrams among all lanthanides, comprising only two energy levels: the ground state $^2F_{7/2}$ and the excited state $^2F_{5/2}$, separated by approximately 10,000 cm$^{-1}$ [31–34]. As a result of an electric interaction of Yb$^{3+}$ ions with host material ions, each of these levels is split into Stark sublevels, with the $^2F_{7/2}$ state exhibiting four Stark components (numbered sequentially from 1 to 4) and the $^2F_{5/2}$ state exhibiting three (numbered 5 to 7) (Figure 2a). Consequently, the emission spectra of phosphors doped with Yb$^{3+}$ typically consist of a single emission band centered at around 1000 nm, dominated by transitions associated with the radiative depopulation of the 5 Stark level of the $^2F_{5/2}$ state[31–33]. The energies of the individual Stark components and the magnitude of their splitting depend on the symmetry of the host material[32,33]. Thus, the thermally induced phase transition in Na$_3$Sc$_2$(PO$_4$)$_3$:Yb$^{3+}$ is expected to alter the shape of the Yb$^{3+}$ emission spectrum. A comparison of the emission spectra of Na$_3$Sc$_2$(PO$_4$)$_3$:Yb$^{3+}$ measured at 83 K and 400 K - representative of the LT and HT phases of Na$_3$Sc$_2$(PO$_4$)$_3$:Yb$^{3+}$, respectively - reveals these differences (Figure 2b). In the case of the LT phase of Na$_3$Sc$_2$(PO$_4$)$_3$:Yb$^{3+}$, distinct Stark lines are observed in the spectrum, and transitions associated with the 6 Stark level are notably present. This unusual for Yb$^{3+}$ ions behaviour results from thermalization of 6 level from 5, caused by limited dispersion of heat induced by the excitation beam[35]. In contrast, the emission spectrum of the HT phase displays broader Stark lines that are not clearly resolved due to their thermal broadening and spectral overlap. Significantly, the spectral positions of the emission lines are shifted in respect to the LT ones, reflecting changes in the strength of Stark splitting for the excited $^2F_{5/2}$ level. This effect is most pronounced for the 5→1 transition, where the energy decreases from 10,331 cm$^{-1}$ in the LT phase to 10,132 cm$^{-1}$ in the HT phase. These structural phase transition-induced changes in the emission spectrum suggest potential applicability of Na$_3$Sc$_2$(PO$_4$)$_3$:Yb$^{3+}$ for ratiometric luminescence thermometry, which is further explored in this study. A comparative analysis of room-temperature emission spectra for

Na$_3$Sc$_2$(PO$_4$)$_3$:Yb$^{3+}$ with varying Yb$^{3+}$ concentrations reveals two primary effects (Figure 2c). First, a gradual red shift in the emission bands is observed with increasing Yb$^{3+}$ ion concentration. This shift is attributed to the increasing contribution of the HT phase in the samples at higher Yb$^{3+}$ concentrations, resulting from a reduction in the phase transition temperature. Second, an increase in the Yb$^{3+}$ ion concentration leads to spectral broadening of individual Stark lines. This broadening may result from structural changes associated with the phase transition or more efficient light-induced heating of the samples[36]. Higher Yb$^{3+}$ concentrations enhance the absorption of the excitation beam, thereby increasing optical heating. However, a similar effect observed with pulsed excitation suggests that the broadening originates primarily from the structural phase transition. Analysis of the luminescence kinetics of Yb$^{3+}$ ions indicates that increasing Yb$^{3+}$ ion concentration leads to greater nonexponentiality in the luminescence decay profiles. To facilitate a comparative analysis of the effect of dopant concentration on the depopulation processes of the $^2F_{5/2}$ level, the average lifetime ($\tau_{avr}$) was determined using the methodology described in the Experimental section. A monotonic decrease in $\tau_{avr}$ from approximately 2 ms for 1% Yb$^{3+}$ to 1.20 ms for 20% Yb$^{3+}$ is observed when Yb$^{3+}$ amount rises up (Figure 2d). The implications of this trend are analysed in detail later in this paper.

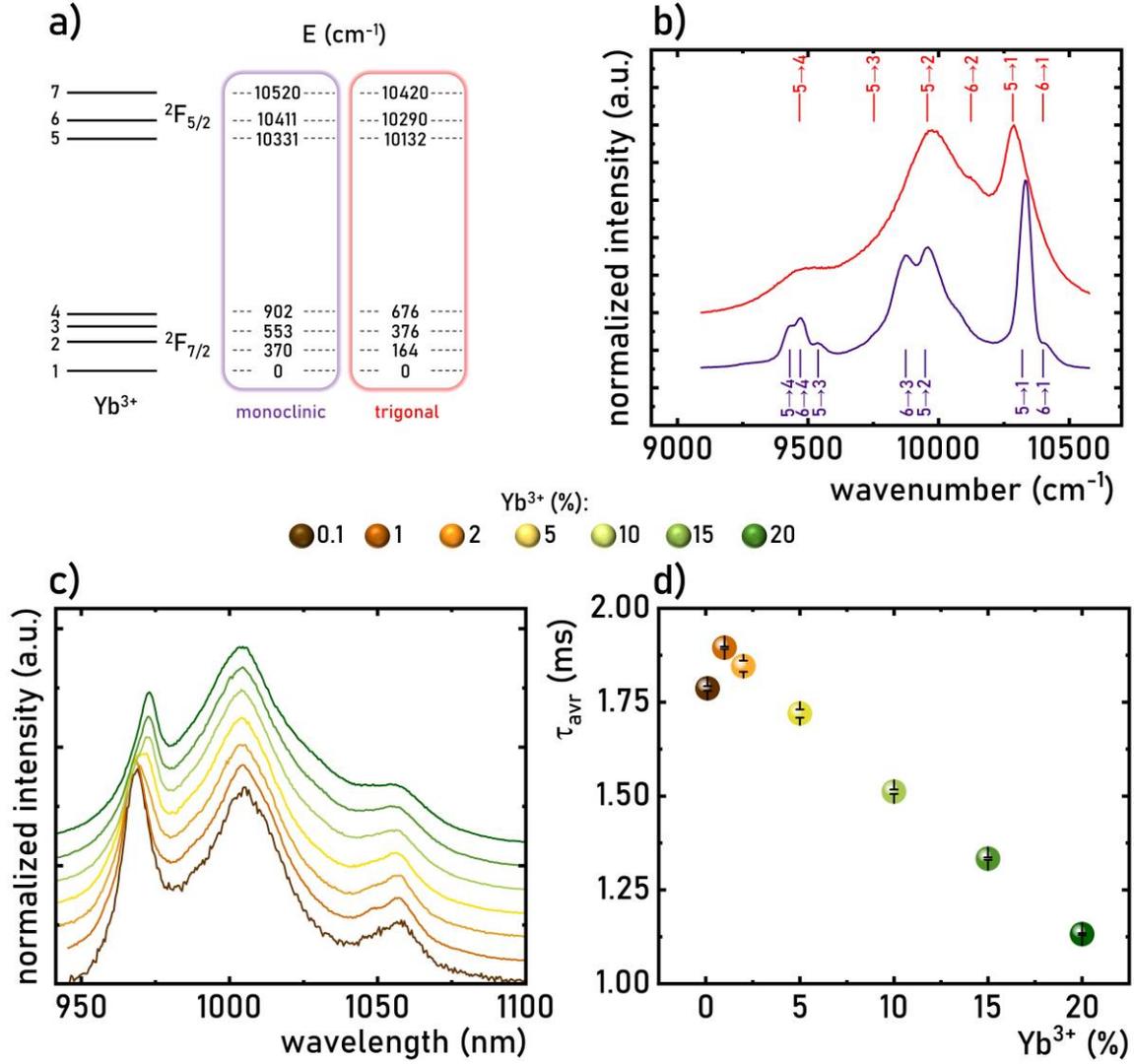

**Figure 2.** Energy diagram of the $Yb^{3+}$ ions with the energies of the Stark levels of $^2F_{5/2}$ and $^2F_{7/2}$ multiplets in monoclinic and trigonal structures of $Na_3Sc_2(PO_4)_3$:$Yb^{3+}$- a); comparison of emission spectra of $Na_3Sc_2(PO_4)_3$:1%$Yb^{3+}$ measured at 83 K (purple line) and 403 K (red line) corresponding to monoclinic and trigonal structures of $Na_3Sc_2(PO_4)_3$:$Yb^{3+}$, respectively -b); the comparison of room temperature emission spectra of $Na_3Sc_2(PO_4)_3$:$Yb^{3+}$ for different $Yb^{3+}$ ions concentration -c) and the influence of the $Yb^{3+}$ ions concentration on the $\tau_{avr}$ of $Na_3Sc_2(PO_4)_3$:$Yb^{3+}$ measured at 83 K ($\lambda_{exc}$=940 nm; $\lambda_{em}$=980 nm) -d).

The thermally induced structural transition in $Na_3Sc_2(PO_4)_3$:$Yb^{3+}$, as described above, modulates the shape of the emission spectrum of $Yb^{3+}$ ions associated with the $^2F_{5/2}\rightarrow{}^2F_{7/2}$ electronic transition. As visualized in normalized luminescence maps, an increase in temperature leads to a progressive decrease in the intensity of Stark components associated with the monoclinic phase of $Na_3Sc_2(PO_4)_3:Yb^{3+}$ and a concurrent increase in those corresponding to the trigonal phase (Figure 3a, Figures S31-37). For $Na_3Sc_2(PO_4)_3:1\%Yb^{3+}$, these changes are particularly pronounced around the phase transition temperature of approximately 350 K. Consequently, the intensity ratio between the monoclinic and trigonal phases can be used as a thermometric parameter. Since the Stark lines transitions of both phases spectrally overlap, it is necessary to deconvolute the luminescence spectra to obtain the luminescence intensity ratio (*LIR*) for signals derived from each phase (Figure S38). This parameter allows for remote temperature measurement.

$$LIR_1 = \frac{\int\limits_{10,243 cm-1}^{10,373 cm-1} \left(^2F_{5/2} \rightarrow\ ^2F_{7/2}\right)_{trigonal} d\nu}{\int\limits_{10,243 cm-1}^{10,407 cm-1} \left(^2F_{5/2} \rightarrow\ ^2F_{7/2}\right)_{monoclinic} d\nu} \qquad (3)$$

As observed, an increase in temperature results in a gradual decrease in $LIR_1$ up to approximately 310 K, beyond which its sharp drop is evident (Figure 3b). Above 380 K, $LIR_1$ stabilizes, with no significant changes at higher temperatures. The concentration of $Yb^{3+}$ ions significantly influences the dynamics of these thermal changes. As $Yb^{3+}$ ion concentration increases, the initial $LIR_1$ value (measured at 83 K) decreases. This effect can be explained by growing contribution of trigonal phase of $Na_3Sc_2(PO_4)_3:Yb^{3+}$ in the sample. For a high concentration of $Yb^{3+}$ ions (e.g., 20%), a temperature increase results in a small but monotonic rise in $LIR_1$ over the entire analyzed thermal range. This behavior can be attributed to several factors. First, the spectral broadening of emission bands arises from slight differences in the site symmetry of crystallographic positions occupied by $Yb^{3+}$ ions. High dopant concentrations often promote structural modifications, as observed in stoichiometric phosphors[37,38]. Band broadening causes 'spectral leaking' between emission intensities of bands originating from both phases. Additionally, higher $Yb^{3+}$ ion concentrations reduce the average distance between neighbouring ions, facilitating interionic energy transfer processes such as energy migration and reabsorption[39] These effects distort the intensity of resonance transitions (e.g., the 5→1 transition of $Yb^{3+}$), leading to lower $LIR_1$ values.

While the deconvolution of $Yb^{3+}$ luminescence band, especially in the case of low dopant concentrations, enables remote temperature sensing, it complicates the readout process. To address this, a ratiometric technique was proposed, using the ratio of integrated luminescence signals from two spectral bands. This approach simplifies temperature measurement by eliminating the need for deconvolution and enhances the thermal variability of $LIR_2$:

$$LIR_2 = \frac{\int_{965nm}^{969nm} {}^2F_{5/2} \to {}^2F_{7/2} d\lambda}{\int_{1030nm}^{1035nm} {}^2F_{5/2} \to {}^2F_{7/2} d\lambda} \quad (4)$$

Spectral ranges used for $LIR_2$ calculations were subjectively selected to maximize the thermal variation of $LIR_2$ (Figure 3c). In the case of 0.1% $Yb^{3+}$, $LIR_2$ decreases from an initial value (obtained at 83 K) of 4.5 to approximately 2 at 300 K, followed by a rapid decline beyond this temperature (Figure 3c). Above 380 K the stabilization of its value was observed and further temperature increase does not induced changes. An increase in $Yb^{3+}$ concentration reduces the initial $LIR_2$ value at 83 K. However, qualitatively, the thermal variation of $LIR_2$ is similar across concentrations. Quantitative thermal variation in $LIR_1$ and $LIR_2$ can be analyzed by calculating the relative thermal sensitivity ($S_R$) defined as:

$$S_R = \frac{1}{LIR} \frac{\Delta LIR}{\Delta T} \cdot 100\% \quad (5)$$

where ΔLIR corresponds to change of LIR for ΔT change in temperature. In the case of $LIR_1$, maximum $S_R$ values are observed around 350 K, coinciding with the structural phase transition temperature (Figure 3d). The highest $S_R$=0.5% K$^{-1}$, was recorded for 0.1% Yb$^{3+}$. Increasing Yb$^{3+}$ concentration causes monotonical reduction of $S_R$ up to 0.1% K$^{-1}$ for 15% Yb$^{3+}$ (Figure 3e). The temperature corresponding to $S_{Rmax}$ also decreases with increasing Yb$^{3+}$ content, correlating well with the reduction in phase transition temperature (Figure 3f). On the other hand, the greater thermal variability of $LIR_2$, in respect to $LIR_1$, is reflected in higher $S_R$ values (Figure 3g). The maximum $S_R$=1.48% K$^{-1}$, was observed at 340 K for 0.1% Yb$^{3+}$. Similar to $LIR_1$ $S_R$ decreases monotonically with increasing Yb$^{3+}$ concentration. However, for $LIR_2$, no clear correlation between Yb$^{3+}$ concentration and the temperature corresponding to $S_{Rmax}$ was observed (Figure 3h). This may be attributed to the spectral range used to calculate $LIR_2$, which overlaps signals from both phases and gains intensity with band broadening at higher Yb$^{3+}$ concentrations (Figure 3i). It is noteworthy that the thermal range of $LIR_1$ and $LIR_2$ values with $S_R$>0.1% K$^{-1}$ is relatively narrow, spanning only 50-60 K, even for 0.1% Yb$^{3+}$. However the use of $LIR_2$ provides much wider thermal range in which $S_R$>0.1% K$^{-1}$ which is another advantage of this approach.

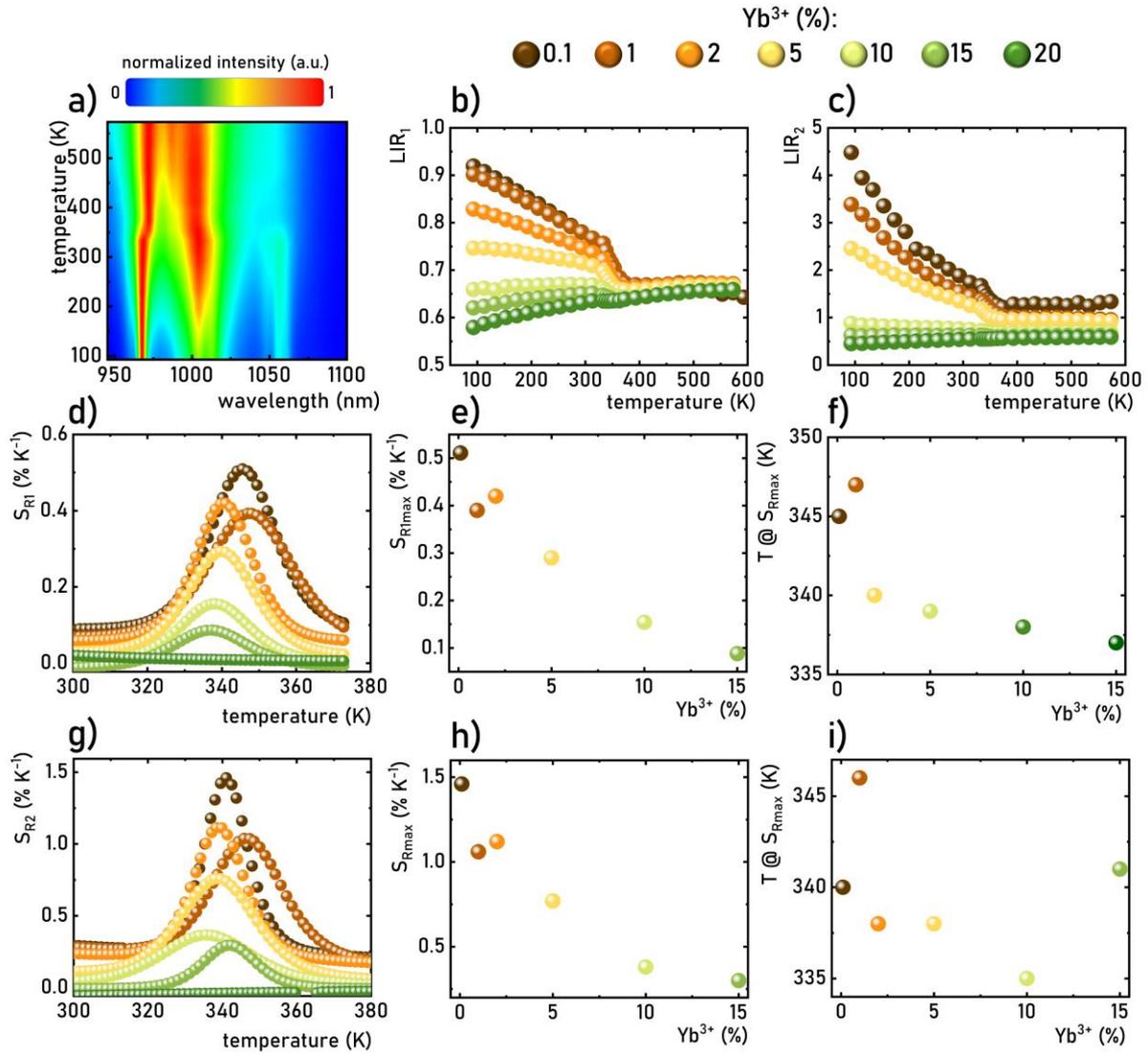

**Figure 3.** Luminescence thermal map of normalized emission spectra of $Na_3Sc_2(PO_4)_3$:1%$Yb^{3+}$-a); thermal dependence of $LIR_1$ -b) and $LIR_2$ -c) for different concentration of $Yb^{3+}$ ions; thermal dependence of $S_R$ corresponding to $LIR_1$; and the influence of $Yb^{3+}$ concentration of the $S_{Rmax}$ -e) and temperature at which $S_{Rmax}$ was obtained- f) for $LIR_1$; thermal dependence of $S_R$ for $LIR_2$ -g) and the influence of $Yb^{3+}$ concentration on corresponding $S_{Rmax}$-h) and temperature at which $S_{Rmax}$ was obtained-i).

The change in the local symmetry of the crystallographic site occupied by $Yb^{3+}$ ions, resulting from the structural phase transition described above, also influences luminescence kinetics (Figure 4a. Figures S39-45). The $Yb^{3+}$ ions are especially favourable to trace these changes due to their simple energy level structure and the high energy separation between levels

(approximately 10,000 cm$^{-1}$), which reduces the probability of nonradiative multiphonon depopulation processes of the excited state, even in host materials of high phonon energies. Consequently, changes in the lifetime of the $^2F_{5/2}$ level associated with the structural transition can be attributed only to variations in the probability of radiative depopulation of the excited state. As shown, increasing the temperature leads to a shortening of the $^2F_{5/2}$ lifetime while maintaining the exponential nature of the luminescence decay profile (Figure 4a). Analysis of the thermal dependence of $\tau_{avr}$ reveals that for temperatures below 300 K, the lifetime remains relatively stable with minimal changes (Figure 4b). However, above 300 K, a sharp shortening of $\tau_{avr}$ is observed, reaching a plateau around 350 K. Similar behavior is observed for all Yb$^{3+}$ ion concentrations below 10%. At higher Yb$^{3+}$ concentrations, a thermal elongation of $\tau_{avr}$ occurs above 400 K. This thermal effect may be attributed to thermally activated energy diffusion between neighbouring $^2F_{5/2}$ levels, which increases the time between energy absorption and photon emission. The reduced interionic distance between dopants at higher Yb$^{3+}$ concentrations facilitates this diffusion process, an effect commonly reported for Yb$^{3+}$-doped materials. The temperature at which this process becomes significant (~400 K) suggests an activation energy for the diffusion process of approximately $k_BT$=250 cm$^{-1}$, which almost corresponds to the energy gap between levels 5 and 7, potentially indicating the involvement of level 7 in this process. The initial $\tau_{avr}$ value at 83 K increases slightly from 1.7 ms for 0.1% Yb$^{3+}$ to 1.78 ms for 2% Yb$^{3+}$. Further increases in dopant ion concentration result in a linear shortening of $\tau_{avr}$. This trend is consistent for $\tau_{avr}$ values immediately following the structural phase transition. Assuming the phase transition temperature ($T_{pt}$) is defined as the arithmetic mean of $\tau_{avr}$ values for the low-temperature (LT) and high-temperature (HT) phases of Na$_3$Sc$_2$(PO$_4$)$_3$:Yb$^{3+}$ (represented by filled circles in Figure 4c), an almost linear correlation between $T_{pt}$ and Yb$^{3+}$ ion concentration is observed (Figure 4d). The rapid change in $\tau_{avr}$ within the temperature range corresponding to the phase transition is reflected in the relative sensitivity

($S_R$) values for lifetime-based luminescence thermometers, which peak in the 315-360 K range (Figure 4e). The maximum $S_R$ value of 1.2% K$^{-1}$ was recorded for Na$_3$Sc$_2$(PO$_4$)$_3$:1%Yb$^{3+}$. An enlargement of the Yb$^{3+}$ concentration leads to a successive reduction in $S_{Rmax}$, reaching 0.2% K$^{-1}$ for Na$_3$Sc$_2$(PO$_4$)$_3$:15%Yb$^{3+}$ (Figure 4f).

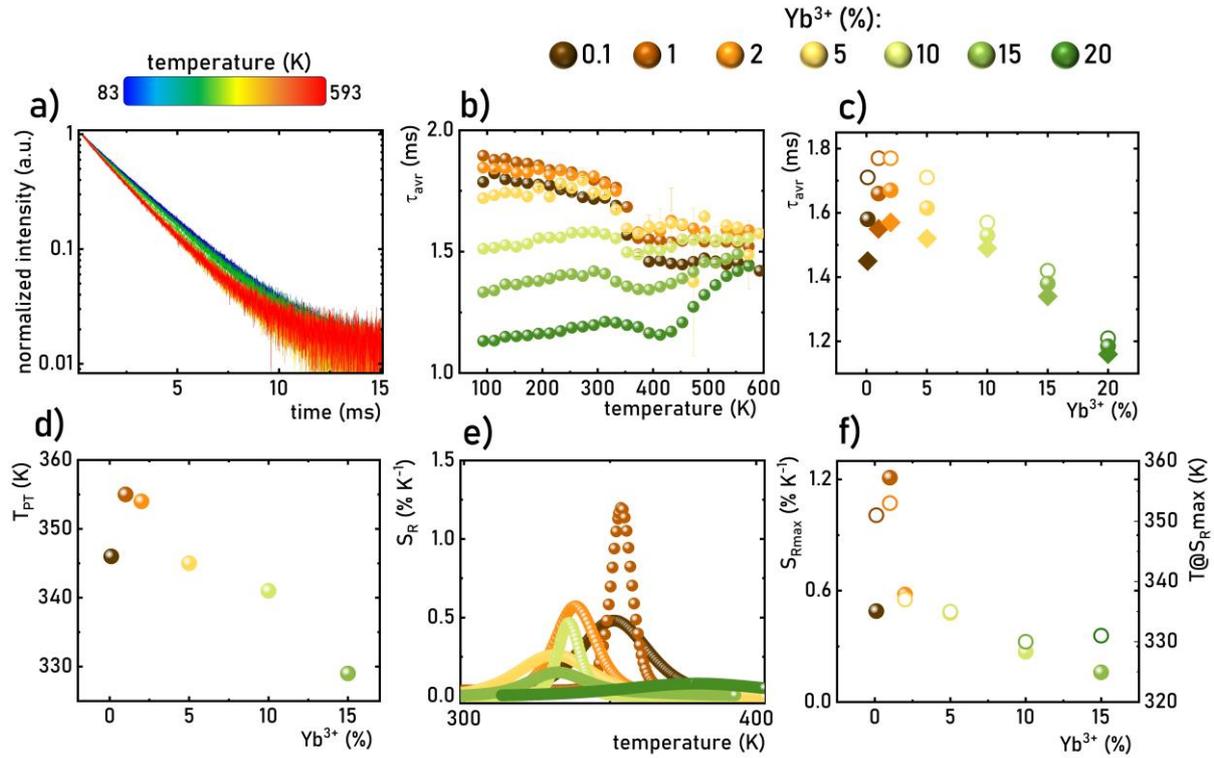

**Figure 4.** Luminescence decay profile of $^2F_{5/2}$ state of Yb$^{3+}$ ions in Na$_3$Sc$_2$(PO$_4$)$_3$:1%Yb$^{3+}$ measured as a function of temperature ($\lambda_{exc}$=940 nm; $\lambda_{em}$=980 nm) -a); thermal dependence of $\tau_{avr}$ for different concentration of Yb$^{3+}$ ions in Na$_3$Sc$_2$(PO$_4$)$_3$:Yb$^{3+}$-b); the influence of Yb$^{3+}$ concentration on $\tau_{avr}$ of low temperature (squares) and high temperature (open circles) phases of Na$_3$Sc$_2$(PO$_4$)$_3$:Yb$^{3+}$ and the mean $\tau_{avr}$ (closed circles) -c); the influence of the Yb$^{3+}$ concentration on the phase transition temperature of Na$_3$Sc$_2$(PO$_4$)$_3$:Yb$^{3+}$ -d); thermal dependence of $S_R$ of the lifetime-based luminescence thermometry -e); the influence of the Yb$^{3+}$ concentration on the $S_{Rmax}$ and temperature at which $S_{Rmax}$ was observed-f).

**Conclusions**

In this study, the structural and luminescence properties of $Na_3Sc_2(PO_4)_3$:$Yb^{3+}$ were investigated as a function of temperature. It was demonstrated that an increase in temperature above 350 K induces a structural transition from monoclinic to trigonal phases. However, increasing the concentration of $Yb^{3+}$ ions lowers the phase transition temperature due to the difference in ionic radii between $Sc^{3+}$ and $Yb^{3+}$, reducing the transition temperature from 345 K for $Na_3Sc_2(PO_4)_3$:0.1%$Yb^{3+}$ to 335 K for $Na_3Sc_2(PO_4)_3$:15%$Yb^{3+}$. This structural transition significantly impacts the spectroscopic properties of $Yb^{3+}$ ions. The increase in symmetry alters the energies of the Stark levels of the excited $^2F_{5/2}$ state and the ground $^2F_{7/2}$ state. The progressive disappearance of Stark lines associated with the monoclinic phase and the concurrent increase in those from the trigonal phase, as temperature rises, enables the development of a ratiometric temperature sensor based on $Na_3Sc_2(PO_4)_3$:$Yb^{3+}$. By deconvoluting the emission spectra of $Yb^{3+}$ ions, it is possible to separate the spectral contributions from each crystallographic phase. The proposed sensor achieves a maximum sensitivity of $S_R$=0.5 %$K^{-1}$ at 345 K. An increase in $Yb^{3+}$ concentration monotonically decreases both the sensitivity and the phase transition temperature, shifting the thermal operating range of the luminescence thermometer. However, a simplified approach that calculates the luminescence intensity ratio of $Yb^{3+}$ ions across two spectral ranges proved not only easier to implement but also resulted in significantly higher relative sensitivity values ($S_{Rmax}$=1.5 %$K^{-1}$ at 340K for $Na_3Sc_2(PO_4)_3$:0.1%$Yb^{3+}$ ). Structural changes in $Na_3Sc_2(PO_4)_3$:$Yb^{3+}$ also influence the luminescence kinetics of the $^2F_{5/2}$ level. Due to the high energy separation between the ground and excited levels of $Yb^{3+}$ ions, the observed rapid shortening of $\tau_{avr}$ above the phase transition temperature can be attributed to changes in the radiative depopulation probability of the $^2F_{5/2}$ state. Utilizing this effect to develop a lifetime-based luminescence thermometer in $Na_3Sc_2(PO_4)_3$:$Yb^{3+}$ yielded sensitivity values ($S_R$=1.2%$K^{-1}$ at 355 K for $Na_3Sc_2(PO_4)_3$:1%$Yb^{3+}$

). The presented research introduces a new paradigm in the design of single-material ratiometric luminescence thermometers. By leveraging structural phase transitions in inorganic materials, it is possible to create luminescence thermometers with high relative sensitivity.

**Acknowledgements**

This work was supported by National Science Center (NCN) Poland under project no. DEC-UMO-2022/45/B/ST5/01629.